\documentclass{moriond}

\usepackage[usenames,dvipsnames]{color}

\usepackage{epsfig}
\usepackage{subfigure}
\usepackage{cite}
\usepackage{amsmath}
\usepackage{mathrsfs}
\usepackage{amssymb}
\usepackage{slashed}

\def\Journal#1#2#3#4{{#1} {\bf #2}, #3 (#4)}

\def\NPB{{\em Nucl. Phys.} B}
\def\PLB{{\em Phys. Lett.}  B}

\def\PRD{{\em Phys. Rev.} D}

\newcommand{\U}{\mathbf{U}}
\newcommand{\T}{\mathbf{T}}
\newcommand{\V}{\mathbf{V}}

\newcommand{\BBd}{B_{\mu\nu}}
\newcommand{\BBu}{B^{\mu\nu}}
\newcommand{\WWd}{W_{\mu\nu}}

\newcommand{\LL}{\mathscr{L}}
\newcommand{\F}{\mathcal{F}}
\renewcommand{\P}{\mathcal{P}}
\renewcommand{\O}{\mathcal{O}}

\DeclareMathOperator{\Tr}{Tr}
\newcommand{\tr}{\Tr}
\DeclareMathOperator{\diag}{diag}
\renewcommand{\to}{\rightarrow}
\newcommand{\de}{\partial}
\newcommand{\nn}{\nonumber}
\newcommand{\hc}{\text{h.c.}}

\newcommand{\e}{\varepsilon}

\newcommand{\Photo}{}

\begin{document}
\vspace*{4cm}
\title{SIGNATURES OF DYNAMICAL SCALARS}

\author{ I.~BRIVIO }

\address{Instituto de F\'isica Te\'orica UAM/CSIC and Departamento de F\'isica Te\'orica,\\
Universidad Aut\'onoma de Madrid, Cantoblanco, 28049 Madrid, Spain}

\maketitle\abstracts{
Effective Lagrangians represent an important, model independent tool for studying physics beyond the Standard Model, via its impact on electroweak scale observables. In particular, two different effective descriptions may be appropriate, depending on how the electroweak symmetry breaking is realized at high energies: a linear effective Lagrangian is best suited in presence of linear dynamics, while a non-linear -chiral- one is more pertinent for scenarios featuring a non-linear realization.
In this talk I will present a few examples of low-energy signals that differentiate the phenomenology of the two descriptions, thus providing a powerful insight into the nature of the Higgs boson.
}

\section{Motivation}
The discovery~\cite{Aad:2012tfa,Chatrchyan:2012ufa} at the LHC of a scalar resonance compatible with the Brout-Englert-Higgs boson (from here on just ``Higgs boson'', for brevity) represents the ultimate experimental proof supporting the Standard Model (SM) of fundamental interactions.
  At the same time, new physics is still expected to exist around the TeV scale in order to cure the theoretical inconsistencies that affect the SM and especially its scalar sector. From this point of view, the observation of the Higgs boson provides a new, unequaled window to shed light on the dynamics of spontaneous electroweak (EW) symmetry breaking (EWSB).

In particular, the viable UV completions of the SM can be classified into two main categories, depending on whether they rely on a linear  or on a non-linear  implementation of EWSB. In the former scenario the Higgs typically appears as an elementary particle, while in the latter framework it arises naturally as a ``dynamical'' -composite- state.

At energies around the EW scale, the impact of beyond-Standard Model (BSM) physics of either class can be described in a model-independent way by means of a Lorentz and gauge-invariant effective Lagrangian. More specifically, a linear effective Lagrangian is pertinent in scenarios with linearly realized EWSB, such as supersymmetric models, where the Higgs particle belongs to an $SU(2)_L$ doublet~$\Phi$ and the new physics scale is $\Lambda\gg v$, being $v$ the EW vacuum expectation value (vev). The linear effective expansion contains operators weighted by inverse powers of the cutoff scale $\Lambda$ and the leading corrections to the SM Lagrangian have then canonical mass dimension $d=6$~\cite{Buchmuller:1985jz,Grzadkowski:2010es,Giudice:2007fh,Low:2009di}.
In dynamical Higgs scenarios, on the other hand, the Higgs particle is a composite field which happens to be a pseudo-goldstone boson (GB) of a spontaneously broken global symmetry.
As a consequence, the most suitable effective Lagrangian for this scenario is a non-linear~\cite{PhysRev.177.2247} or ``chiral" one: a derivative expansion as befits the Goldstone boson dynamics. A typical example of this scenario are composite Higgs models~\cite{Kaplan:1983fs,Kaplan:1983sm,Banks:1984gj,Georgi:1984ef,Georgi:1984af,Dugan:1984hq,Agashe:2004rs,Contino:2006qr,Gripaios:2009pe,Marzocca:2012zn}.

The effective linear and chiral Lagrangians with a light Higgs are in general different: the latter is indeed more general, and contains the former as a special limit which can only be obtained imposing specific constraints by hand~\cite{Brivio:2013pma} or assuming  peculiar dynamics at high energies~\cite{Alonso:2014wta}.
In this talk I present a few examples of signals that differentiate the two scenarios and that can thus provide powerful insights to the origin of the EWSB mechanism.

\section{The effective non-linear Lagrangian for a light Higgs}
The effective low-energy chiral Lagrangian for a light Higgs is
written in terms of the SM fermions and gauge bosons and of two scalar fields:
the SM GBs are described by a dimensionless unitary matrix~\cite{Appelquist:1980vg,Longhitano:1980iz,Longhitano:1980tm,Feruglio:1992wf,Appelquist:1993ka}
$
\U(x)=e^{i\sigma_a \pi^a(x)/v}$, $\U(x) \rightarrow L\, \U(x) R^\dagger\,,
$
with $L,R$ denoting respectively the $SU(2)_{L,R}$ global transformations of the scalar potential. 
The Higgs boson is represented by the singlet field $h$ and its couplings are encoded in generic functions
\begin{equation}\label{F}
 \F_i(h) = 1+2a_i\frac{h}{v}+b_i\frac{h^2}{v^2}+\dots
\end{equation}
that lack any $SU(2)_L$ structure: as often pointed
out (e.g. refs.~\cite{Grinstein:2007iv,Contino:2010mh}), the resulting effective Lagrangian can
describe many setups including that for a light SM singlet isoscalar.
 
 In a phenomenological approach, the effective non-linear Lagrangian for a light Higgs can be written as
$\LL_\text{chiral} = \LL_0 + \Delta \LL$,
where the leading order $\LL_0$ is the SM Lagrangian and $\Delta\LL$ describes any deviation from the SM due to the presence of strong-interacting new physics above the EW scale.
The former term reads then
\begin{equation}\label{LL0}
\begin{split}
\LL_0 =& \frac{1}{2} (\de_\mu h)(\de^\mu h) -\frac{1}{4}\WWd^a W^{a\mu\nu}-\frac{1}{4}\BBd\BBu-\frac{1}{4} G^a_{\mu\nu}G^{a\mu\nu}- V (h)\\
 &-\frac{(v+h)^2}{4}\tr[\V_\mu\V^\mu]+ i\bar{Q}\slashed{D}Q+i\bar{L}\slashed{D}L\\
 &-\frac{v+  h}{\sqrt2}\left(\bar{Q}_L\U \mathbf{Y}_Q Q_R+\hc\right)-\frac{v+ h}{\sqrt2}\left(\bar{L}_L\U \mathbf{Y}_L L_R+\hc\right)\,,
\end{split}
\end{equation} 
where \mbox{$\V_\mu\equiv \left(D_\mu\U\right)\U^\dagger$} ($\T\equiv\U\sigma_3\U^\dag$) is the vector (scalar) chiral field transforming in the adjoint of $SU(2)_L$.
The covariant derivative is 
\begin{equation}
D_\mu \U(x) \equiv \de_\mu \U(x) +igW_{\mu}(x)\U(x) - 
                      \frac{ig'}{2} B_\mu(x) \U(x)\sigma_3 \, , 
\end{equation}
with $W_\mu\equiv W_{\mu}^a(x)\sigma_a/2$ and $B_\mu$ denoting the
$SU(2)_L$ and $U(1)_Y$ gauge bosons, respectively. In eq.~\eqref{LL0},
the first line describes the $h$ and gauge boson kinetic terms, and
the effective scalar potential $V(h)$. The second line describes the $W$ and $Z$
masses and their interactions with $h$, as well as the kinetic terms
for GBs and fermions.  Finally, the third line corresponds to the
Yukawa-like interactions written in the fermionic mass eigenstate
basis. A compact notation for the
right-handed fields has been adopted, gathering them into
doublets 
$Q_R$ and $L_R$. $\mathbf{Y}_Q\equiv\diag\left(Y_U,\, Y_D\right)$ and $\mathbf{Y}_L\equiv  \diag\left(Y_\nu,\, Y_L\right)$ are two $6\times6$ block-diagonal matrices containing the
usual Yukawa couplings.

The term $\Delta\LL$ includes all the effective operators with up to four derivatives allowed by Lorentz and gauge symmetries. 
In the bosonic (pure gauge, pure Higgs and gauge-$h$ operators), CP even sector, to which we restrict in this talk\footnote{The bosonic CP odd sector is analyzed in~\cite{Gavela:2014vra}, while a complete basis comprehensive of both bosonic and fermionic operators has been proposed in~\cite{Buchalla:2013rka}. 
}, it can be decomposed as
\begin{align}
\Delta\LL= & c_B\P_B(h) + c_W\P_W(h)+c_G\P_G(h) +c_C \P_C(h) + c_T \P_T(h)+c_H \P_H(h)+\nn\\
   &+c_{\Box H} \P_{\Box H}(h)+\sum_{i=1}^{26} c_i\P_i(h)
\label{DeltaL}
\end{align}
where $c_i$ are model-dependent coefficients, and the explicit form of the operators $\P_i(h)$ can be read from ref.~\cite{Brivio:2013pma}.

\section{Phenomenology: signatures of non-linearity}
In order to identify the phenomenological signatures that differentiate the linear form the non-linear EFTs, it is useful to compare the chiral set of operators (eq.\eqref{DeltaL}) with a complete basis of dimension six, bosonic, CP even linear operators. Here we choose the so-called HISZ linear basis~\cite{Hagiwara:1993ck,Hagiwara:1996kf} and we report the main results of the exhaustive analysis performed in refs.~\cite{Brivio:2013pma,Brivio:2014pfa}.

Exploiting the correspondence  $ \Phi=(v+h)/\sqrt2 \,\U \left( 0\,\;1\right)^T$, it is possible to identify two main categories of discriminating effects:
 \begin{enumerate}
  \item Some couplings  are predicted to be correlated in the linear parameterization, but receive contributions from independent operators in the non-linear description.
  For example, the linear term $\O_B=(ig'/2)\BBd D^\mu\Phi^\dag D^\nu\Phi$ is set in correspondence with the combination of two non-linear terms:
  \begin{equation}\label{OB}\small
   \O_B \:\to\:\frac{ig'v^2}{16}\BBd\big[\tr(\T[\V^\mu,\V^\nu]) \F_{2}(h)+2 \tr(\T\V^\mu) \de^\nu\F_{4}(h)\big]= \frac{v^2}{16}\big[\P_2(h)+2\P_4(h)\big]\,,
  \end{equation} 
  where $\P_2(h)$ contributes to the TGCs usually dubbed $\kappa_Z$ and $\kappa_\gamma$, while $\P_4(h)$ introduces the anomalous HVV vertices $A_{\mu\nu}Z^\mu\de^\nu h$ and $Z_{\mu\nu}Z^\mu\de^\nu h$. 
  
  In a linear scenario any departure of one of these couplings from its SM value is expected to be correlated with effects in the other three, since they all receive a contribution from $\O_B$ (obviating for the time being all the other possible operators).  Moreover, the relative magnitude of such contributions is fixed by the structure of the covariant derivative $D_\mu\Phi$.
  In the most general non-linear framework, instead, no such correlation is present: deviations in $\{\kappa_Z,\kappa_\gamma\}$ are parameterized in terms of the coefficient $c_2$, while those in the two anomalous HVV vertices are proportional to $c_4$.  This effect is due to the different gauge representation chosen in the two theories for the Higgs field: in the chiral formalism the Higgs particle $h$ is treated as a gauge singlet, independent of the three SM GBs. As a consequence, this framework lacks the strong link between the couplings of the Higgs and those of the longitudinal gauge bosons, which in the linear realization is imposed by the doublet structure of the field $\Phi$.
  A completely analogous analysis holds for another linear operator, $\O_W=(ig/2)\,D^\mu\Phi^\dag\WWd D^\nu\Phi$, that contributes to the same TGV and HVV vertices as $\P_2(h),\,\P_4(h)$ and corresponds to the chiral operators $\P_3(h)$ and $\P_5(h)$:
  \begin{equation}\small
   \O_W \:\to\: \frac{igv^2}{8}\big[\tr(\WWd [\V^\mu,\V^\nu]) \F_{3}(h)-2 \tr(\WWd\V^\mu) \de^\nu\F_{5}(h)\big] =  \frac{v^2}{8}\big[\P_3(h)-2\P_5(h)\big]\,.
  \end{equation} 
 
  In the event of some anomalous observation in either of the couplings mentioned above, the presence or absence of correlations 
would allow for direct testing of the nature of the Higgs boson. A preliminary global-fit analysis on the four parameters $c_{2-5}$ was presented in ref.~\cite{Brivio:2013pma}.
Analogous (de)correlation effects between couplings with different number of Higgs legs have been discussed in refs.~\cite{Contino2012,Isidori:2013cga}. Finally, a more complex example, that involves the six chiral operators $\P_{\Box h},\,\P_{6-10}$ is analyzed in ref.~\cite{Brivio:2014pfa}.
  
    \item Some couplings appear only at higher order in the linear expansion, i.e. in linear operators of dimension $d\geq 8$, but are allowed as first-order corrections to the SM (i.e. at the four-derivatives level) in the non-linear description.
This kind of signal arises as a consequence of the adimensionality of the $\U(x)$ matrix, which ensures that the GB contributions do not exhibit any scale suppression. This is in contrast with the linear description, where the light $h$ and the three SM GBs are encoded into the scalar doublet $\Phi$, with mass dimension one: in that case any insertion of $\Phi$ pays the price of a suppression factor $1/\Lambda$.
   
  As an example, the operator $\P_{14}(h)=g \e^{\mu\nu\rho\lambda} \tr(\T\V_\mu) \tr(\V_\nu W_{\rho\lambda}) \F_{14}(h)$ contains the anomalous TGC $\e^{\mu\nu\rho\lambda}\de_\mu W^+_\nu W^-_\rho Z_\lambda$, called $g_5^Z$ in the parameterization of~\cite{Hagiwara:1986vm}. This coupling appears only at dimension 8 in the linear expansion. Therefore, if found to be non-zero and comparable in size to other leading corrections to the SM, this effect would represent a smoking gun of non-linearity in the EWSB sector. 
Current limits on $g_5^Z$ are derived from LEP data; however, the LHC has the potential to improve these bounds: the study presented in~\cite{Brivio:2013pma}, based on the kinematical analysis of the process $pp\to W^\pm Z\to\ell'^{\pm}\ell^+\ell^-\slashed{E}_T$, shows that with a luminosity of 300~fb$^{-1}$ at a c.o.m. energy of 14~TeV it is possible to measure $g_5^Z$ at a level of precision comparable to that of the current constraints on dimension 6 linear operators.

 \end{enumerate}

\vspace*{-2mm}

\section*{Acknowledgments}
My work is supported by an ESR contract of the European Union network FP7 ITN INVISIBLES (Marie Curie Actions, PITN-GA-2011-289442). I also acknowledge partial support of the Spanish MINECO’s “Centro de Excelencia Severo Ochoa” Programme under grant SEV-2012-0249.
and I thank the organizers of the Moriond conference for the kind invitation and for their efforts in organizing this enjoyable meeting.

\end{document}